\documentclass[usenatbib]{emulateapj}

\shortauthors{Cowan \& Agol}
\shorttitle{Thermal Phase Variations of Eccentric Exoplanets}
\begin{document}

\title{A Model for Thermal Phase Variations of Circular and Eccentric Exoplanets}

\author{Nicolas B. Cowan\altaffilmark{1,2},
	 Eric Agol\altaffilmark{2},
	 }

\altaffiltext{1}{CIERA Fellow, Northwestern University, Dearborn Observatory, 2131 Tech Drive, Evanston, IL 60208}
\altaffiltext{2}{Astronomy Department, University of Washington,
   Box 351580, Seattle, WA  98195\\
email: n-cowan@northwestern.edu}

\begin{abstract}
We present a semi-analytic model atmosphere for close-in exoplanets that captures the essential physics of phase curves: orbital and viewing geometry, advection, and re-radiation. 
We calibrate the model with the well-characterized transiting planet, HD~189733b, then compute light curves for seven of the most eccentric transiting planets: Gl~436b, HAT-P-2b, HAT-P-11b, HD~17156b, HD~80606b, WASP-17b, XO-3b. We present phase variations for a variety of different radiative times and wind speeds.
In the limit of instant re-radiation, the light curve morphology is entirely dictated by the planet's eccentricity and argument of pericenter: the light curve maximum leads or trails the eclipse depending on whether the planet is receding from or approaching the star at superior conjunction, respectively. For a planet with non-zero radiative timescales, the phase peak occurs early for super-rotating winds, and late for sub-rotating winds.
We find that for a circular orbit, the timing of the phase variation maximum with respect to superior conjunction indicates the direction of the dominant winds, but cannot break the degeneracy between wind speed and radiative time. For circular planets the phase minimum occurs half an orbit away from the phase maximum ---despite the fact that the coolest longitudes are always near the dawn terminator--- and therefore does not convey any additional information.
In general, increasing the advective frequency or the radiative time has the effect of reducing the peak-to-trough amplitude of phase variations, but there are interesting exceptions to these trends.
Lastly, eccentric planets with orbital periods significantly longer than their radiative time exhibit ``ringing'' whereby the hot spot generated at periastron rotates in and out of view. The existence of ringing makes it possible to directly measure the wind speed (the frequency of the ringing) and the radiative time constant (the damping of the ringing). 
\end{abstract}


\section{Introduction}
Efforts to model the atmospheres of exoplanets fall broadly into two categories: 1) radiative transfer models that make predictions about the planet's spectral characteristics after making assumptions about the bulk energy transport in the atmosphere \citep[numerical simulations include][while Hansen 2008 developed a series of analytic radiative transfer models, and Madhusudhan \& Seager 2009 introduced a Monte Carlo approach for fitting models to observed spectra]{Iro_2005, Seager_2005, Barman_2005, Burrows_2006, Fortney_2008}; 2) hydrodynamic models that make predictions about energy recirculation given assumptions about radiative transfer. The hydrodynamic models run with a relatively simple treatment of radiative transfer may then be post-processed with more detailed radiative transfer \citep[e.g.,][]{Burrows_2010}.  Roughly in increasing order of sophistication, the hydrodynamic models consist of solving the: shallow water equations \citep{Langton_2007, Langton_2008}, equivalent barotropic equations \citep{Cho_2003, Cho_2008, Rauscher_2008}, primitive equations \citep{Showman_2002, Cooper_2005, Showman_2008, Showman_2009, Menou_2009, Rauscher_2009, Thrastarson_2010}, Euler's equations \citep{Burkert_2005, Dobbs_2008}, or the full Navier Stokes equation \citep{Dobbs-Dixon_2010}. Recently, \cite{Showman_2009} and \cite{Lewis_2010} have combined approaches 1 and 2 by simultaneously solving the radiative transfer \emph{and} hydrodynamical equations. With the exceptions of the hydrodynamical models of \cite{Langton_2008}, \cite{Lewis_2010} and \cite{Beaulieu_2010}, as well as the radiative transfer models of \cite{Iro_2010}, however, these theoretical efforts have been directed towards understanding exoplanets on circular orbits, for which the planet's power budget is constant.  

Despite the difficulties in numerically modeling exoplanet atmospheres, the physical parameters that govern their large-scale behavior are thought to be: the radiative timescale (how quickly a parcel of gas radiates away its thermal energy), and the advective timescale (how long it takes for a parcel of gas to move from the planet's day-side to its night-side). In detail, not every joule of energy is absorbed at the same depth (let alone re-radiated from that depth) and different parcels of gas will take different amounts of time to move from the day-side to the night-side. Nevertheless, lacking detailed observations of these planets, such a simple approach complements more detailed simulations by indicating ---at very little computational cost--- which ingredients are necessary to properly explain observed behavior. 

The radiative timescale is primarily determined by the depths at which incident energy from the host star is being absorbed and emitted (which are not necessarily the same). Radiative transfer models suggest that this occurs at pressures of $10^{-3}$--$10^{0}$~bars, where the radiative time scales are roughly $10^4$--$10^5$~seconds \citep{Seager_2005, Iro_2005, Fortney_2005, Barman_2005, Burrows_2006}. Advection timescales are a function of planetary size and wind speed. Different hydrodynamic simulations put the wind speeds at roughly a tenth of the sound speed at the slow end, and transonic at the high end: $\sim10^2$--$10^3$~m/s \citep{Showman_2002,Cho_2003,Burkert_2005,Cooper_2005,Cooper_2006,Langton_2007, Thrastarson_2010}.  Such wind speeds lead to advective times of $\sim10^5$~s for a Jupiter-sized planet\footnote{Unlike the radiative timescale, which depends sensitively on depth, the wind velocity decreases by only a factor of 3 between 2.5~mbar and 20~bar \citep{Cooper_2005}.}. Since the radiative and advective timescales are of the same magnitude, it is not immediately clear how efficient energy recirculation should be for short-period exoplanets. Indeed, hydrodynamic simulations performed by different groups (see above) do not always agree.

Fortunately, there is a growing body of observations that constrain theoretical models for short-period planets. 
Observations of secondary eclipses in exoplanetary systems have made 
it possible to measure the integrated day-side flux of short-period exoplanets \citep[e.g.,][]{Charbonneau_2005, Deming_2005}. Observations of phase variations have made it possible to measure the brightness of planets as a function of \emph{local time} \citep{Harrington_2006, Cowan_2007, Knutson_2007a, Knutson_2009a, Snellen_2009a, Borucki_2009, Knutson_2009b, Crossfield_2010}. Ideally, there would be analytic models to bridge the gap between the detailed simulations and these cutting-edge phase observations. We presented a summary and analysis of the photometric observations of 24 transiting planets in \cite{Cowan_2010}. We concluded that short-period planets exhibit a wide range of recirculation efficiencies but are all consistent with low Bond albedos ($A<0.35$).  The model used in that paper, however, was too simple to predict the \emph{shape} of phase variations, only their amplitude.

Phase function mapping \citep{Cowan_2008} is a largely model-independent technique, but it rests on the assumption that large-scale weather patterns on the planet do not shift with respect to the sub-stellar point. There is theoretical evidence that hot Jupiter atmospheres reach a steady-state diurnal heating pattern over a wide range of pressures  that include the likely mid-IR photosphere \citep[2.5~mbar -- 19.6~bar,][]{Cooper_2005}. But this lack of variability may be an artifact of initial conditions \citep[][]{Thrastarson_2010}. Nevertheless, repeat observations of HD~189733b with \emph{Spitzer} indicate that the 8~micron day-side variability is less than 2.7\% and night-side variability is less than 17\% \citep[][]{Agol_2009, Agol_2010}. Thus, the steady-state hypothesis seems to be in good shape for the large-scale climate of planets on circular orbits, but it is impossible for this condition to hold true on planets with eccentric orbits\footnote{Unless, of course, the radiative timescale is longer than an orbital period, in which case the planet would not exhibit \emph{any} phase variations (see Equation~\ref{T_flat}).}.

Here we develop an analytic energy transport model---complementary to hydrodynamic models---which offers important insight into the bulk energy flow in the atmospheres of hot Jupiters, and its impact on the primary observable: thermal phase variations. In \S~2 we introduce our analytic model and the associated differential equation (DE). We describe general solutions and useful limits of the DE in \S~3. In \S~4 we explain how we compute disc-integrated light curves. We compute sample light curves in \S~5, discuss our model assumptions and how they likely impact our results in \S~6, and state our conclusions in \S~7.

\section{Energy Transport Model}
In this section, we describe a one-dimensional (longitudinal) analytic model of the heating and cooling of a planet. For now we simply state our simplifying assumptions and develop the equations governing our model. In \S~6 we discuss these assumptions explicitly. We assume that all of the energy is reflected, absorbed and emitted by a single layer, which should be taken to approximate the planet's photosphere, or emitting surface.  

The temperature and ---for planets in edge-on orbits--- the visibility of parcels of gas is greatest near the equator. Since we are interested only in disc-integrated light curves, we are principally concerned with the motion of the atmosphere near the equator. We assume that the equatorial jet stream has a constant angular velocity with latitude \citep[i.e. solid body rotation. See also][]{Iro_2005}. We define this constant effective rotation period for the planet, $\omega_{\rm rot}$, in an inertial frame. Note that from the perspective of a parcel of gas, it doesn't matter whether this motion is a result of actual rotation of the planet, winds, or both. Winds are usually defined with respect to a rotating core. It is difficult (currently impossible) to determine the rotation rate of a gaseous exoplanet's core, $\omega_{\rm core}$. Observationally, the only accessible quantity is the sum of the planet's internal rotation and the winds, which we refer to as the planet's effective rotation: $\omega_{\rm rot} = \omega_{\rm core} + \omega_{\rm wind}$. Note that the effective rotation of the planet can be trivially converted to an effective rotational period, $P_{\rm rot}=2\pi/\omega_{\rm rot}$, or an equatorial parcel velocity in an inertial frame, $v_{\rm gas} = R_{p}\omega_{\rm rot}$. 

For the energy budget of the parcel of gas, what matters is not the rotation frequency but rather the related \emph{advective} frequency, $\omega_{\rm adv}(t)=\omega_{\rm rot} - \omega_{\rm orb}(t)$, where $\omega_{\rm orb}$ is the angular velocity of the planet about its host star, which varies with time for an eccentric orbit. The advective frequency is a measure of how often the parcel of gas is subjected to stellar forcing. A planet with a tidally locked \emph{atmosphere}\footnote{As opposed to a tidally locked \emph{core}.} has $\omega_{\rm adv}=0$; super-rotating winds (in the same sense as the planet's orbital motion) have $\omega_{\rm adv}>0$; sub-rotating winds have $\omega_{\rm adv}<0$. The orbital angular velocity is a maximum at periastron: $\omega_{\rm max} = (2\pi/P)(1-e)^{-3/2}(1+e)^{1/2}$, where $P$ is the planet's orbital period. Note that $\omega_{\rm max}$ is not the same as the pseudo-synchronous rotational frequency (see \S~6 for discussion on this). In this study, we will often express $\omega_{\rm rot}$ in units of $\omega_{\rm max}$ since this immediately indicates whether the planet's atmosphere is super-rotating or trailing when its day-side is receiving the largest incident flux.

\subsection{Energy Transport Differential Equation}
The flux incident on a planet changes as a function of time if the planet's orbit is not circular:
\begin{equation}
F(t) = \sigma T_{\rm eff}^{4} \left(\frac{R_{*}}{r(t)}\right)^{2},
\end{equation}
where $\sigma$ is the Stefan-Boltzmann constant, $T_{\rm eff}$ and $R_{*}$ are the star's effective temperature and radius, and $r(t)$ is the planet--star distance. 

We assume that parcels of gas move only in the E--W direction, not latitudinally, and we adopt the Lagrangian approach of tracking an individual parcel of gas and its temperature. The flux absorbed by the parcel is $(1-A)F(t)\sin\theta\max(\cos\Phi(t), 0)$, where $A$ is the planet's Bond albedo (which we assume to be constant), $\theta$ is the co-latitude of the parcel ($\theta=0$ at the north pole, $\pi/2$ at the equator, and $\pi$ at the south pole), and $\Phi$ is the angle from the sub-stellar longitude (a.k.a. local time: $\Phi=-\pi/2$ at the dawn terminator; $\Phi=0$ at the substellar meridian; $\Phi=\pi/2$ at the dusk terminator; $\Phi=\pi$ at the antistellar meridian).

We further assume that all of the heat absorbed by the parcel goes into increasing its temperature, rather than doing mechanical work. The warmed gas then radiates as an ideal blackbody with temperature $T$ (This is simply Newtonian cooling, a scheme used in many of the hydrodynamic simulations listed in \S~1).  The differential equation governing the temperature of this parcel of gas as a function of time is:
\begin{equation} \label{de}
\frac{dT}{dt} = \frac{1}{c_{h}}\left( (1-A)F(t)\sin\theta\max(\cos\Phi(t), 0) - \sigma T^{4}\right),
\end{equation}
where $c_{h} = \rho C_{P} H$ is constant everywhere on the planet, provided that the mass-density, $\rho$, specific heat capacity, $C_{P}$, and the thickness of the parcel, $H$ are planet-wide constants. Since all the incident flux is absorbed (and eventually re-radiated) in a single layer, one must take this layer to be many atmospheric scale-heights thick: including the optical and infrared photospheres. Note that even if the dynamical layer of the atmosphere is a dozen scale heights thick, it is still much smaller than the planetary radius.

Setting the left side of Equation~\ref{de} to zero, we define the equilibrium temperature for a parcel of gas:
\begin{equation}
T_{\rm eq}(\Phi, \theta, t)=\left( \frac{(1-A)F(t)\sin\theta\max(\cos\Phi(t), 0) }{\sigma}\right)^{\frac{1}{4}}. 
\end{equation}
A parcel cools if its temperature is greater than $T_{\rm eq}$; it heats up if its temperature is below $T_{\rm eq}$. The temperature extrema of the parcel (and any inflection points) thus occur when its temperature reaches $T_{\rm eq}$. Note that latitude enters the differential equation in the same way as Bond albedo.  This means that locating a parcel farther from the equator is precisely equivalent to increasing its Bond albedo: $(1-A) \leftrightarrow \sin\theta$.

The energy budget of an eccentric planet is dictated by what happens near periastron. Noting that the time variability of $T_{\rm eq}$ is contained in $\Phi(t)$ and $F(t)$, we define the fiducial temperature along the sub-stellar meridian at periastron \citep[see also][]{Hansen_2008, Cowan_2010}:
\begin{equation}
T_{0}= T_{\rm eff} (1-A)^{1/4} \sin^{1/4}\theta \sqrt{\frac{R_{*}}{a(1-e)}}.  
\end{equation}
Following \cite{Iro_2005} and \cite{Seager_2005}, we define the radiative time constant in the vicinity of the photosphere as 
\begin{equation}
\tau_{\rm rad} = c_{h}/\sigma T_{0}^{3}.
\end{equation}
 
We define the dimensionless temperature, $\tilde{T} = T/T_{0}$, and dimensionless time\footnote{In the $\tau_{\rm rad}\to 0$ limit, the dimensionless time is poorly defined, but Equation~\ref{de} can be trivially solved, as shown in \S~\ref{epsilon_limits}.}, $\tilde{t} = t/\tau_{\rm rad}$, and rewrite Equation~\ref{de} as
\begin{equation} \label{dimensionless}
\frac{d \tilde{T}}{d\tilde{t}} = \tilde{I}- \tilde{T}^{4},
\end{equation}
where $\tilde{I} = \max(\cos\Phi(t), 0) (a(1-e)/r(t))^2$ is the normalized intensity of radiation experienced by the parcel of gas. 

\section{Energy Transport Solutions}
Equation~\ref{dimensionless} is a first-order, non-linear ordinary differential equation (D.E.) and does not have an analytic solution on the planet's day-side. On the planet's night-side, $\tilde{I} \equiv 0$, and the equation admits an analytic solution: 
\begin{equation}
\tilde{T} = \tilde{T}_{\rm dusk} \left( \frac{1}{3(\tilde{t}-\tilde{t}_{\rm dusk}) \tilde{T}_{\rm dusk}^{3} +1} \right)^{1/3},
\end{equation}
where $\tilde{T}_{\rm dusk}$ and $\tilde{t}_{\rm dusk}$ are the dimensionless temperature and time at sunset.

We can ---and eventually do--- solve the D.E. numerically, but in the remainder of this section we consider simplifying limits that offer insight into the model behavior. Readers primarily interested in the calculated light curves may skip ahead to \S~4. 

\subsection{The Limits of Low and High $\tau_{\rm rad}$} \label{epsilon_limits}
The limiting case of $\tau_{\rm rad}=0$ corresponds to a planet where all the flux is re-radiated from the day-side, regardless of how rapidly the gas parcels are moving. Every parcel of gas in such a case is always right at its equilibrium temperature: $T(\Phi, \theta, t) = T_{\rm eq}(\Phi, \theta, t)$, given by Equation~3.

Given our model assumptions, the $\tau_{\rm rad} \gg P_{\rm rot}$ limit corresponds to a planet that emits radiation uniformly at all longitudes. For a planet on an eccentric orbit, the limit comes in two varieties: if $P_{\rm rot} \ll \tau_{\rm rad} \ll P$, the planet instantly re-radiates all incident energy, but uniformly in longitude. This may be thought of as a planet with strong zonal winds, leading to efficient day--night heat circulation:
\begin{equation} 
T(\theta, t) = T_{\rm eff}\left(\frac{R_{*}}{r(t)}\right)^{1/2} \left(\frac{1}{\pi} \right)^{1/4} \sin^{1/4}\theta (1-A)^{1/4}.
\end{equation}
 If $P_{\rm rot} \ll P \ll \tau_{\rm rad}$, the radiative time is so long that one probably cannot neglect latitudinal (meridional) transport; it is expedient to assume perfect latitudinal transport. The planet will emit uniformly everywhere on the planet and at the same rate all the time: 
\begin{equation} \label{T_flat}
T = T_{\rm eff}\left(\frac{R_{*}}{2a}\right)^{1/2} (1-A)^{1/4} (1-e^{2})^{-1/8}.
\end{equation}

\subsection{Circular Orbit Limit}
In the special case of circular orbits (the majority of short-period exoplanets, for good reason), there are many useful analytic approximations that we can make to simplify the problem. Furthermore, the simpler circular case offers intuition into the behavior of our model. In the circular limit, Equation~\ref{dimensionless} can be rewritten as:
\begin{equation} \label{circular_de}
{d \tilde{T} \over d\Phi} = {1 \over \epsilon} 
\left(\max(\cos\Phi,0) - \tilde{T}^4\right),
\end{equation}
where $\epsilon = \tau_{\rm rad}\omega_{\rm adv}$ is a dimensionless constant quantifying the planet's energy recirculation efficiency\footnote{For simplicity, we consider here only positive $\epsilon$.  That is, super-rotating equatorial jets. Later in the paper we consider both super-rotating and trailing winds.}, and $\Phi = \omega_{\rm adv} t$. 

\begin{figure}[htb]
\includegraphics[width=84mm]{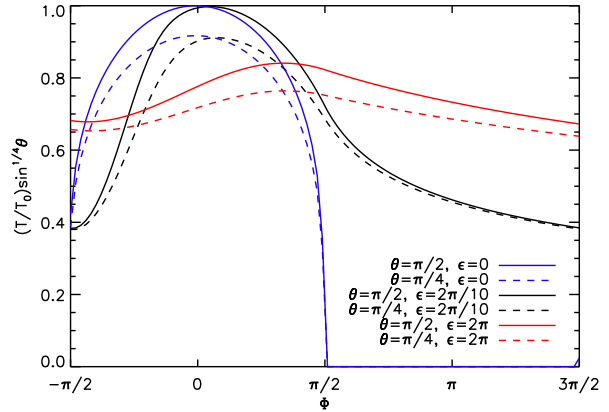}
\caption{Temperature of a parcel of gas in the atmosphere of a short-period planet in a circular orbit. The plot runs from dawn ($\Phi = -\pi/2$) until dusk ($\Phi = \pi/2$), and back to dawn again ($\Phi = 3\pi/2$).}
\label{heating}
\end{figure}

A day in the life of a parcel of gas proceeds as shown in Figure~\ref{heating}. Note that we have included the $\sin^{1/4}\theta$ factor in the y-axis, so that we may plot on the same figure the heating curves for parcels at different latitudes. The two effects of a high $\epsilon$ are 1) a delay in the time of maximum temperature, and  2) higher night-time temperature. Although the observed light curve depends on both the radiative and advective timescales, the heating pattern of a parcel of gas depends only on their ratio, $\epsilon$.  The night-time temperature is largely independent of latitude, $\theta$, but depends sensitively on $\epsilon$.  The maximum temperature reached by a parcel, on the other hand, depends sensitively on its latitude but only weakly on $\epsilon$. Because of the equivalency of latitude and albedo, the the dashed lines in Figure~\ref{heating} can be thought of as the heating curves for equatorial parcels of gas, but with $A \approx 30$\%: albedo has a more important impact on the day-side heating pattern than on the night-side cooling. Finally, the delay between a parcel passing through the sub-stellar longitude and reaching its maximum temperature, $\Phi_{\rm max}$, depends more sensitively on $\epsilon$ than does the maximum temperature reached, $\tilde{T}_{\rm max}$. In the $\epsilon \to \infty$ limit, $\tilde{T} \sin^{1/4}\theta = (1/\pi)^{1/4}$, as one would expect from Equation~\ref{dimensionless}. 

The diurnal heating patterns shown in Figure~\ref{heating} are similar to those measured at the surface of Earth. In both cases parcels of gas move in and out of the sunlight: on Earth this motion is entirely due to the planet's rotation (wind velocities are small compared to rotational velocity), while on a gaseous planet this motion is due to a combination of rotation and zonal winds. Indeed, our model may also be relevant to a rocky planet with a thin atmosphere and rapid rotation, provided that the rotational period is shorter than the lateral heat conduction timescale.  

\begin{figure}[htb]
\includegraphics[width=84mm]{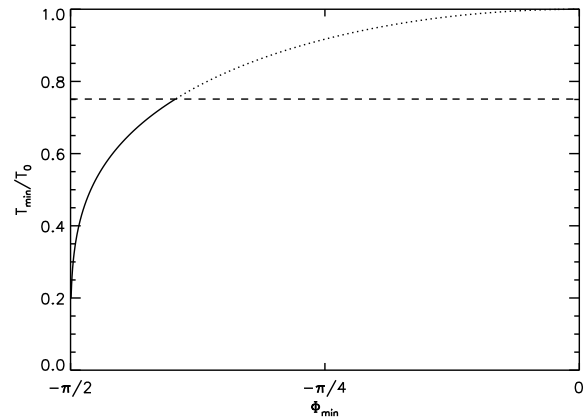}
\caption{The solid line shows the minimum temperature reached by a parcel of gas, $\tilde{T}_{\rm min}$, plotted against the longitude at which the minimum is reached ($\Phi_{\rm min}=-\pi/2$ at the dawn terminator, $\Phi_{\rm min}=0$ at the sub-stellar point). The dashed line is the maximum value of $\tilde{T}_{\rm min}=(1/\pi)^{1/4}$, reached in the $\epsilon \to \infty$ limit. We show the rest of the functional form of Equation~\ref{phi_min_eqn} as a dotted line for completeness.}
\label{phi_min}
\end{figure}

\begin{figure}[htb]
\includegraphics[width=84mm]{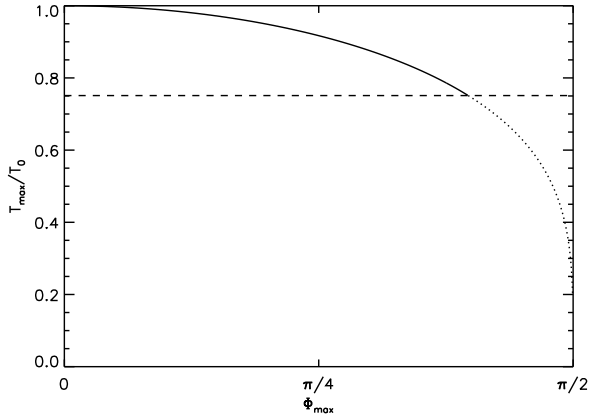}
\caption{The solid line shows the maximum temperature reached by a parcel of gas, $\tilde{T}_{\rm max}$, plotted against the longitude at which the maximum is reached ($\Phi_{\rm max}=0$ at the sub-stellar point, $\Phi_{\rm max}=\pi/2$ at the dusk terminator). The dashed line is the minimum value of $\tilde{T}_{\rm max}=(1/\pi)^{1/4}$, reached in the $\epsilon \to \infty$ limit. We show the rest of the functional form of Equation~\ref{phi_max_eqn} as a dotted line for completeness.}
\label{phi_max}
\end{figure}

For a planet on a circular orbit, we can use the fact that the temperature extrema for a particle of gas occur when it reaches its equilibrium temperature. The temperature extrema are therefore related to their location on the planet by:
\begin{equation} \label{phi_min_eqn}
\tilde{T}_{\rm min} = \cos^{\frac{1}{4}}\Phi_{\rm min}
\end{equation}   
and
\begin{equation} \label{phi_max_eqn}
\tilde{T}_{\rm max} = \cos^{\frac{1}{4}}\Phi_{\rm max},
\end{equation}
where $\Phi_{\rm min}$ and $\Phi_{\rm max}$ are the angles between the sub-stellar meridian and temperature minimum and maximum, respectively. Since $\tilde{T}_{\rm min}$ is typically much less than unity, $\Phi_{\rm min}\approx -\pi/2$ in most situations (see Figure~\ref{phi_min}), while $\tilde{T}_{\rm max}$ is close to unity so small differences in this maximum temperature correspond to significant changes in the phase offset of the maximum, as shown in Figure~\ref{phi_max}.

Note that the the diurnal heating pattern has a longitudinal asymmetry: parcels of gas are heated faster than they cool, so parcels East of the hot-spot tend to be warmer than those West of the hot-spot. This asymmetry manifests itself in the disc-integrated thermal phase curve of the planet: the offset in the peak of the lightcurve tends to be larger than $\Phi_{\rm max}$.

\begin{figure}
\includegraphics[width=84mm]{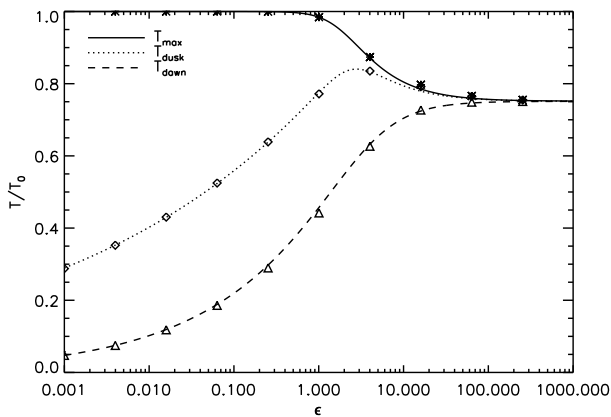}
\caption{Dimensionless temperature, $\tilde{T}=T/T_{0}$ at the hot-spot, at dusk and at dawn, as a function of recirculation efficiency, $\epsilon$. The lines show the analytic expressions presented in the Appendix, while the point symbols show the result of numerically solving Equation~\ref{dimensionless}. Although the analytic approximation does a good job over a variety of $\epsilon$, it is only applicable for planets on circular orbits.}
\label{t_sunset_vs_epsilon}
\end{figure}

In the Appendix we develop analytic approximations in the circular regime for $T_{\rm max}$, $T_{\rm dusk}$ and $T_{\rm dawn}$ (the maximum temperature reached by a parcel, its temperature at the dusk terminator, and at the dawn terminator). Those analytic approximations are compared to the numerical solutions in Figure~\ref{t_sunset_vs_epsilon}. For intermediate recirculation efficiencies, $\epsilon \approx 3$, the sunset temperature has a peak.
This is due to advection of heat downstream, which becomes more efficient as $\epsilon$
increases, heating the sunset terminator.   However, when $\epsilon$ becomes too large,
heat is distributed uniformly along a given longitude, causing the temperature to
decrease again, hence the peak at intermediate $\epsilon$.
 
\section{Applications of Model}
By combining the energy transport model described above with a geometrical model of the planet's motions (both orbital and rotational), we now produce model phase variations for an entire planet. We model the planet as a grid of gas parcels: 40 longitude and 20 latitude grid points. We run the calculation for 3 orbital periods, with 1000 time steps per orbit. We adopt a perfectly edge-on orbital geometry in all cases, but otherwise use the appropriate orbital parameters ($a$, $e$, $\omega$), planetary radius ($R_{p}$), and stellar parameters ($R_{*}$, $T_{\rm eff}$, $\log g$, [Fe/H]). We can run the calculations for arbitrary wavelengths but only present here the light curves for \emph{Spitzer} IRAC channel 4 (8~$\mu$m). Since ours is a one-layer model, the shape of the phase variations is essentialy unchanged if one adopts a different waveband. The only free model parameters are $A$, $\tau_{\rm rad}$, and $\omega_{\rm rot}$; in practice we set $A=0$ since the albedo does not significantly affect the \emph{shape} of the thermal phase variations and it appears that hot Jupiters have low albedos \citep[][and references therein]{Rowe_2008, Cowan_2010}. 
 
All stellar and planetary data are taken from exoplanet.eu, maintained by Jean Schneider; using numbers from exoplanets.org did not perceptibly change our results. When the stellar data are not available, we have assumed typical parameters for the appropriate spectral class, and solar metallicity.  Insofar as we are only concerned with the broadband mid-IR brightnesses of the stars, our results should not depend sensitively on the input stellar parameters. Using the stars' $T_{\rm eff}$, $\log g$ and [Fe/H], we use the PHOENIX/NextGen stellar spectrum grids \citep{Hauschildt_1999} to determine their brightness temperatures at the observed frequencies. For each waveband, we determine the ratio of the stellar flux to the blackbody flux at that grid star's $T_{\rm eff}$.  We then apply this factor to the $T_{\rm eff}$ of the actual observed star.

There are many computational shortcuts that one can use with this analytic model.  The orbits are treated as edge-on and the planetary obliquity is assumed to be zero (this is strictly true for the core; for the atmosphere it simply means that the equatorial jet-stream flows in the East-West direction), so the D.E. need only be solved at one latitude on the day-side (e.g., the equator) and those day-side heating curves are easily adjusted for other latitudes via $T_{0}$. The D.E. can then be solved analytically on the planet's night-side. 

\subsection{Planet-Integrated Properties}
The observed flux ratio depends on a combination of orbital factors ($a$, $e$), planetary factors ($A$, $\tau_{\rm rad}$, $\omega_{\rm rot}$), and viewing geometry ($\omega$ and $\alpha$, the usual phase angle: $\alpha=0$ at eclipse, $\alpha=\pi$ at transit). We show schematically in Figure~5 how we combine the orbital, planetary and viewing factors to obtain disc-integrated thermal light curves.
The top panel simply shows the planet's distance from its host star; the second panel shows the planet's orbital angular velocity; the third panel shows the equilibrium temperature at the sub-stellar point (solid line) and the highest temperature on the model planet (dotted line). The fourth panel shows the total absorbed flux (solid line) and the total emitted flux (dotted line). The fifth panel shows the planet's illuminated fraction, $f = \frac{1}{2}(1+\cos\alpha)$. The bottom panel shows the planet/star flux ratio at 8~$\mu$m as seen from Earth.  

\begin{figure}
\includegraphics[width=84mm]{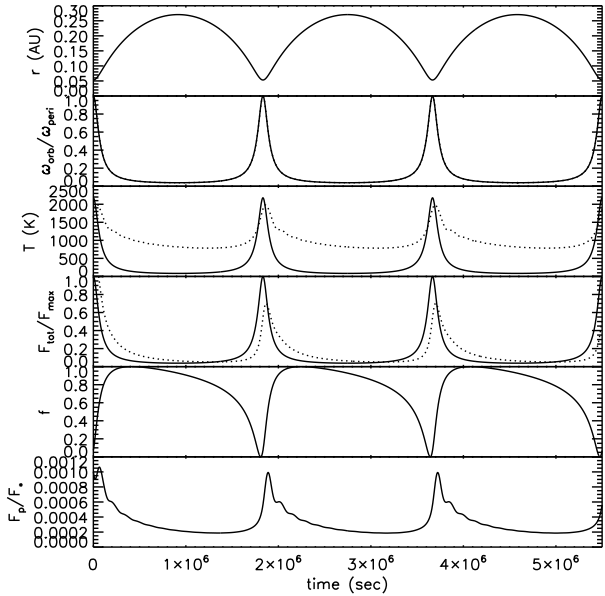}
\caption{Diagnostics for HD~17156b, using a model with $A=0$, $\tau_{\rm rad}=20$~hours, $\omega_{\rm rot}=2\omega_{\rm max}$.}
\label{diagnostics_plot}
\end{figure}

We start the calculations with all of the parcels at $T = T_{0}$, but the planet reaches a periodic equilibrium in a couple e-folding times (a few $\tau_{\rm rad}$). If the planet had no heat capacity, the dotted lines would perfectly track the solid lines in the third and fourth panels of Figure~5. The effect of a non-zero planetary heat capacity (non-zero $\tau_{\rm rad}$) is: the peak-to-trough swings in emitted flux are damped, and the peak in emission lags the peak in absorption.  

A planet on a circular orbit reaches a steady state whereby the temperature of a parcel of gas is always the same at a given latitude, $\theta$, and local time, $\Phi$. In other words, the planet's climate is steady as seen from the host star. This condition cannot occur for an eccentric planet.  Instead, we verify that our calculations satisfy a somewhat weaker steady-state condition: the parcels of gas at the same latitude, local time \emph{and orbital phase} should have the same temperature, orbit after orbit. Since we use radiative timescales much shorter than any of the orbital periods considered here, reaching this steady state is easily achieved by solving the heating curve D.E. for a full three orbits.

The thermal contrast ratio between the planet and its host star amounts to a weighted mean of the brightness of the visible hemisphere of the planet:
\begin{equation}
\frac{F_{p}}{F_{*}} = \frac{1}{\pi B(T_{\rm b})}\left(\frac{R_{p}}{R_{*}} \right)^{2} \oint V(\phi, \theta, t) B(T) d\Omega,
\end{equation}
where $B(T_{\rm b})$ is the star's intensity given its brightness temperature at this wavelength, $B(T)$ is the blackbody intensity at the temperature $T(\phi, \theta, t)$, and $V(\phi, \theta, t)$ is the visibility of a region on the planet \citep[$V=1$ at the sub-observer point, drops as the cosine of the angle from the observer, and is null on the far side of the planet.  For details, see][]{Cowan_2009}.

\section{Model Light Curves}
\begin{figure}[tb]
\includegraphics[width=84mm]{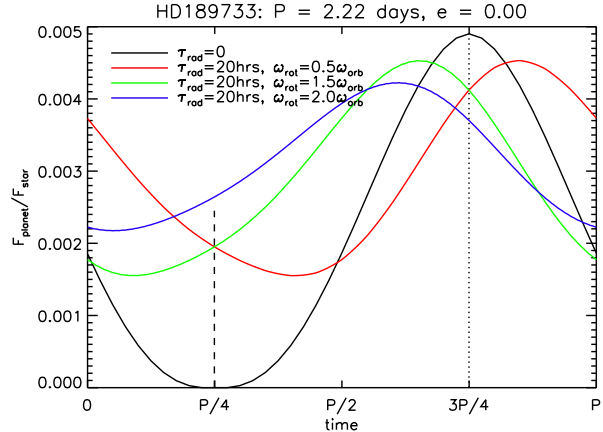}
\caption{The 8~$\mu$m light curve for HD~189733b \citep{Bouchy_2005} for a variety of radiative and advective timescales, and assuming $A=0$. The vertical dashed and dotted lines mark inferior conjunction (transit) and superior conjunction (eclipse), respectively. The transit and eclipses have been removed for clarity. For non-zero $\epsilon$, the min and max of the light curve will not necessarily occur at inferior and superior conjunction. The effective rotation rate of the atmosphere is expressed in units of the orbital frequency, $\omega_{\rm orb}$. The peak of the light curve occurs after superior conjunction for $\omega_{\rm rot} < \omega_{\rm orb}$; it occurs before superior conjunction for $\omega_{\rm rot} > \omega_{\rm orb}$.}
\label{sample_HD189733}
\end{figure}

Sample light curves for a planet on a circular orbit (HD~189733b) are shown in Figure~\ref{sample_HD189733}. The minimum in the light curve occurs half an orbit from the maximum, despite the fact that the minimum temperature reached by a parcel of gas is always near the dawn terminator. 

As discussed in Agol et al. (2010), the 8~$\mu$m observations of HD~189733b (phase variations and eclipse offset) indicate an $\epsilon$ of order unity and super-rotating winds, shown by the blue line in Figure~\ref{sample_HD189733}. Without further information, it is impossible to break the degeneracy between $\tau_{\rm rad}$ and $\omega_{\rm rot}$ \citep[this degeneracy has been pointed out in different terms in][]{Cho_2008}. But as mentioned in \S~1, many hydrodynamical simulations of hot Jupiters indicate trans-sonic winds \citep[eg:]{Cooper_2005, Rauscher_2009}, making for planet-crossing times of order 10$^{5}$~seconds ($=30$~hours). This means that the radiative timescale of HD~189733b's atmosphere is in the tens of hours. For our calculations, we adopt either the $\tau_{\rm rad}=0$ limit, or a fiducial radiative time of 20~hours, and consider three different advective schemes: $\omega_{\rm rot}=0.5\omega_{\rm orb}$ is a trailing wind; it takes parcels of gas two planetary orbits to experience one diurnal cycle. The $\omega_{\rm rot}=1.5\omega_{\rm orb}$ is a slightly super-rotating wind; it again takes parcels of gas two planetary orbits to experience one diurnal cycle. For the fast super-rotating winds ($\omega_{\rm rot}=2.0\omega_{\rm orb}$), parcels of gas experience one diurnal cycle per orbit.   

The sub-rotating ($\omega_{\rm rot}=0.5\omega_{\rm orb}$) and super-rotating ($\omega_{\rm rot}=1.5\omega_{\rm orb}$) cases have opposite phase offsets but identical maxima and minima. At superior conjunction they have identical contrast ratios, so one cannot distinguish between the two models with secondary eclipse depth alone. In detail, of course, the timing and shape of secondary eclipse \emph{is} sensitive to the day-side brightness map of the planet \citep[][]{Williams_2006, Rauscher_2007, Agol_2010}. 

\begin{figure}[tb]
\includegraphics[width=84mm]{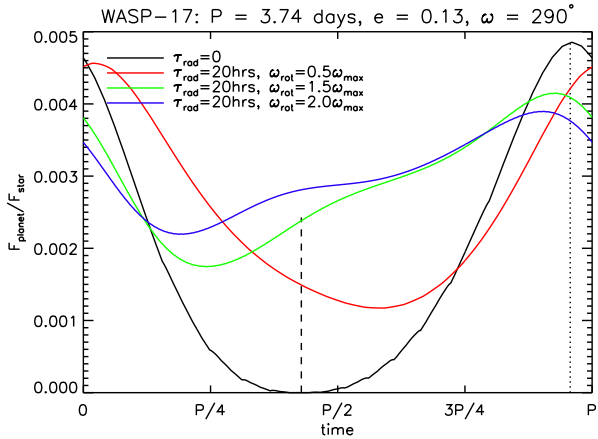}
\caption{The 8~$\mu$m light curve for WASP-17b \citep{Anderson_2010} for a variety of radiative and advective timescales, and assuming $A=0$. The vertical dashed and dotted lines mark inferior conjunction (transit) and superior conjunction (eclipse), respectively. The transit and eclipses have been removed for clarity. The effective rotation rate of the atmosphere is expressed in units of the maximal orbital frequency, $\omega_{\rm max}$.}
\label{sample_WASP-17}
\end{figure}

\begin{figure}[tb]
\includegraphics[width=84mm]{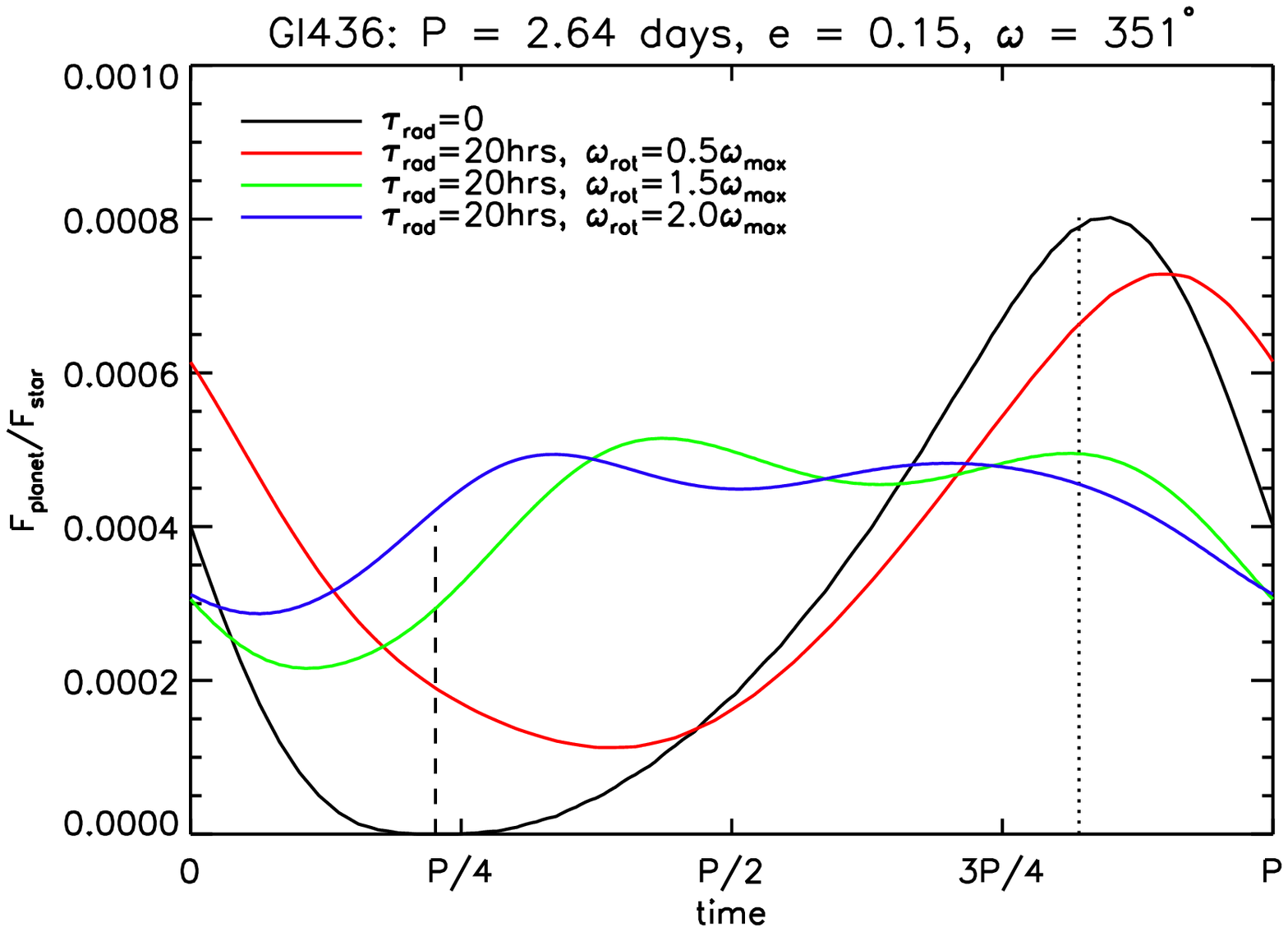}
\caption{The 8~$\mu$m light curve for Gl~436b \citep{Butler_2004} for a variety of radiative and advective timescales, and assuming $A=0$. The vertical dashed and dotted lines mark inferior conjunction (transit) and superior conjunction (eclipse), respectively. The transit and eclipses have been removed for clarity. The effective rotation rate of the atmosphere is expressed in units of the maximal orbital frequency, $\omega_{\rm max}$.}
\label{sample_Gl436}
\end{figure}

\begin{figure}[tb]
\includegraphics[width=84mm]{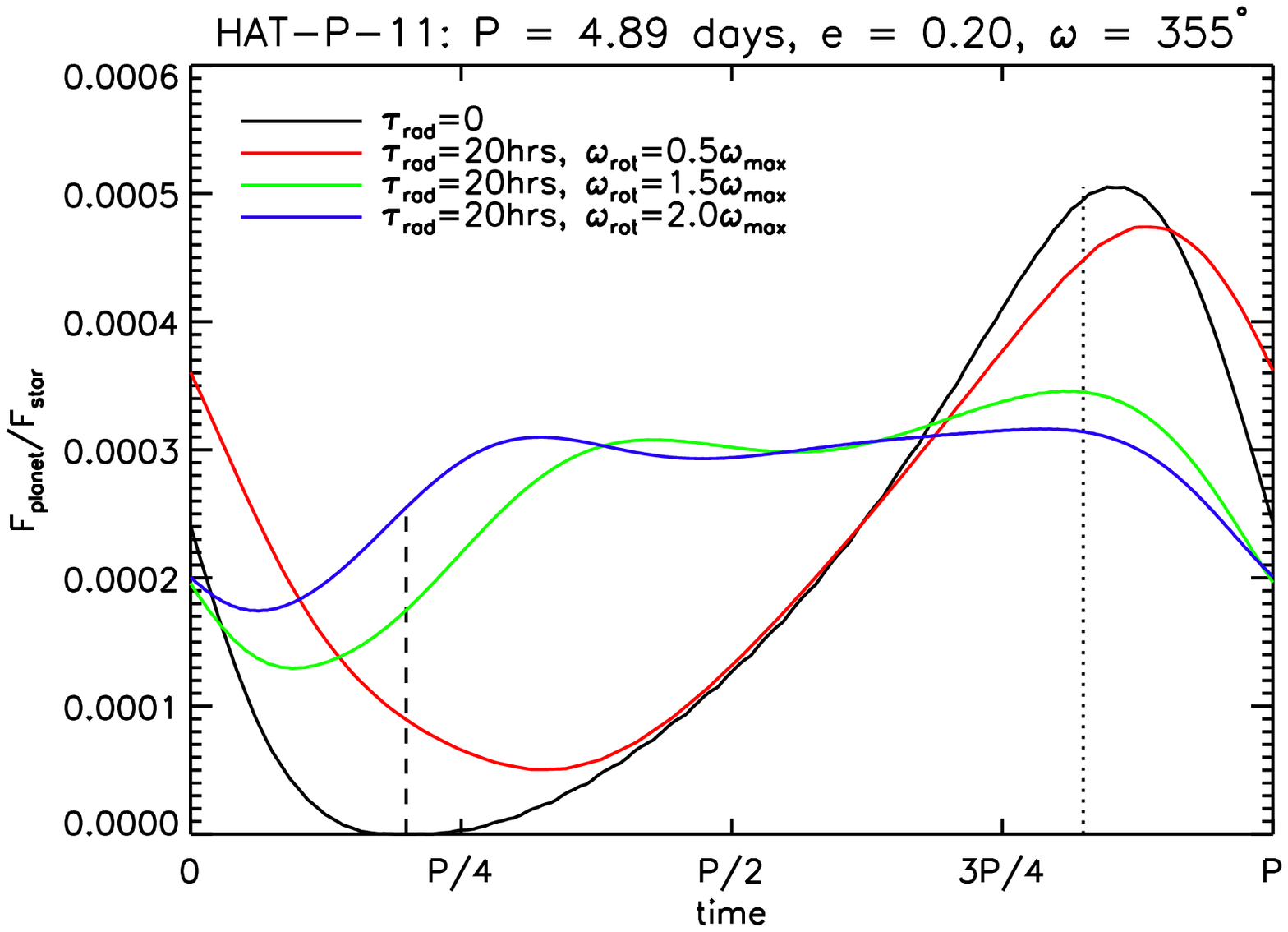}
\caption{The 8~$\mu$m light curve for HAT-P-11b \citep{Bakos_2010} for a variety of radiative and advective timescales, and assuming $A=0$. The vertical dashed and dotted lines mark inferior conjunction (transit) and superior conjunction (eclipse), respectively. The transit and eclipses have been removed for clarity. The effective rotation rate of the atmosphere is expressed in units of the maximal orbital frequency, $\omega_{\rm max}$.}
\label{sample_HAT-P-11}
\end{figure}

\begin{figure}[tb]
\includegraphics[width=84mm]{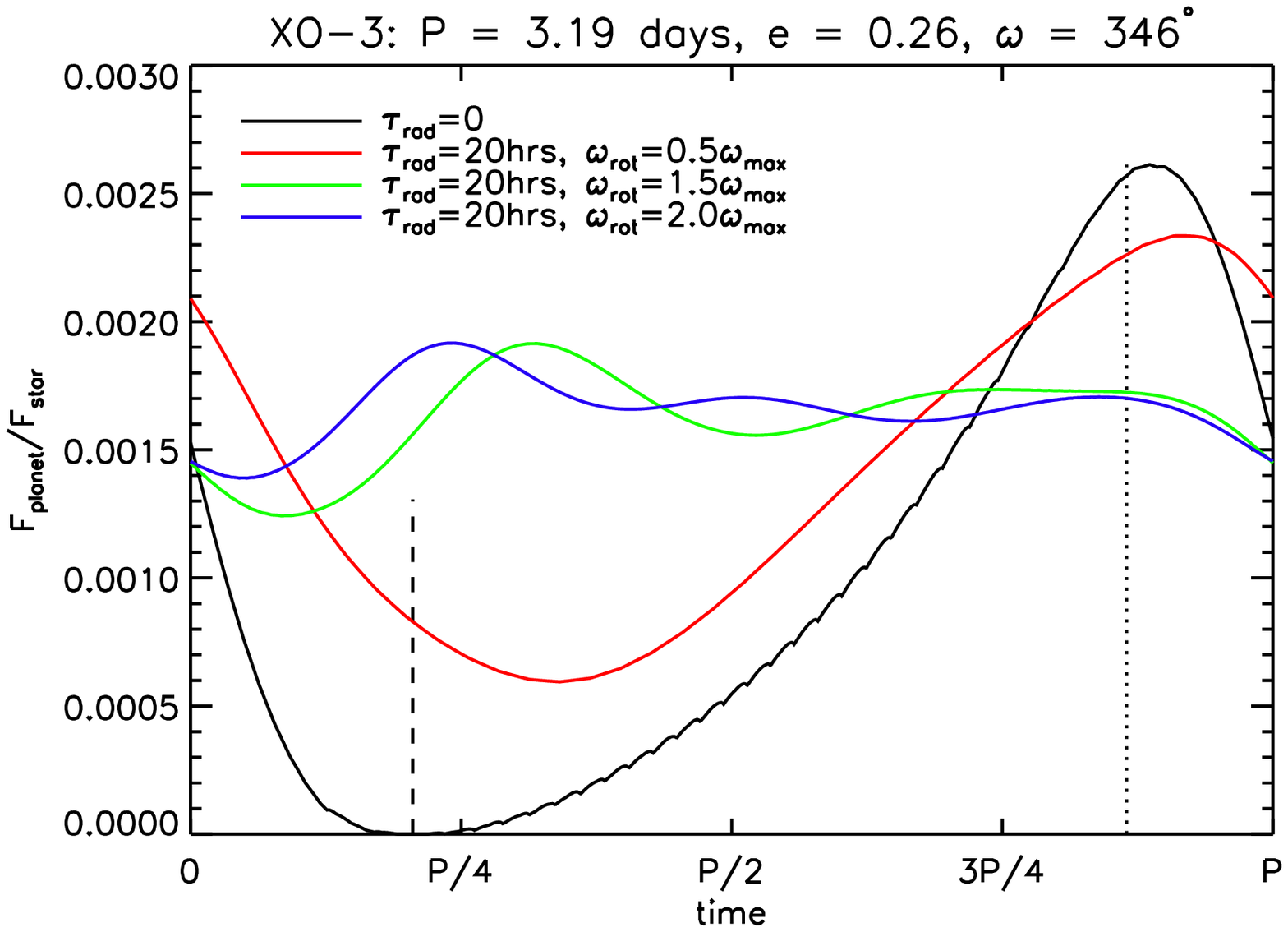}
\caption{The 8~$\mu$m light curve for XO-3b \citep{Johns-Krull_2008} for a variety of radiative and advective timescales, and assuming $A=0$. The vertical dashed and dotted lines mark inferior conjunction (transit) and superior conjunction (eclipse), respectively. The transit and eclipses have been removed for clarity. The effective rotation rate of the atmosphere is expressed in units of the maximal orbital frequency, $\omega_{\rm max}$.}
\label{sample_XO-3}
\end{figure}

\begin{figure}[tb]
\includegraphics[width=84mm]{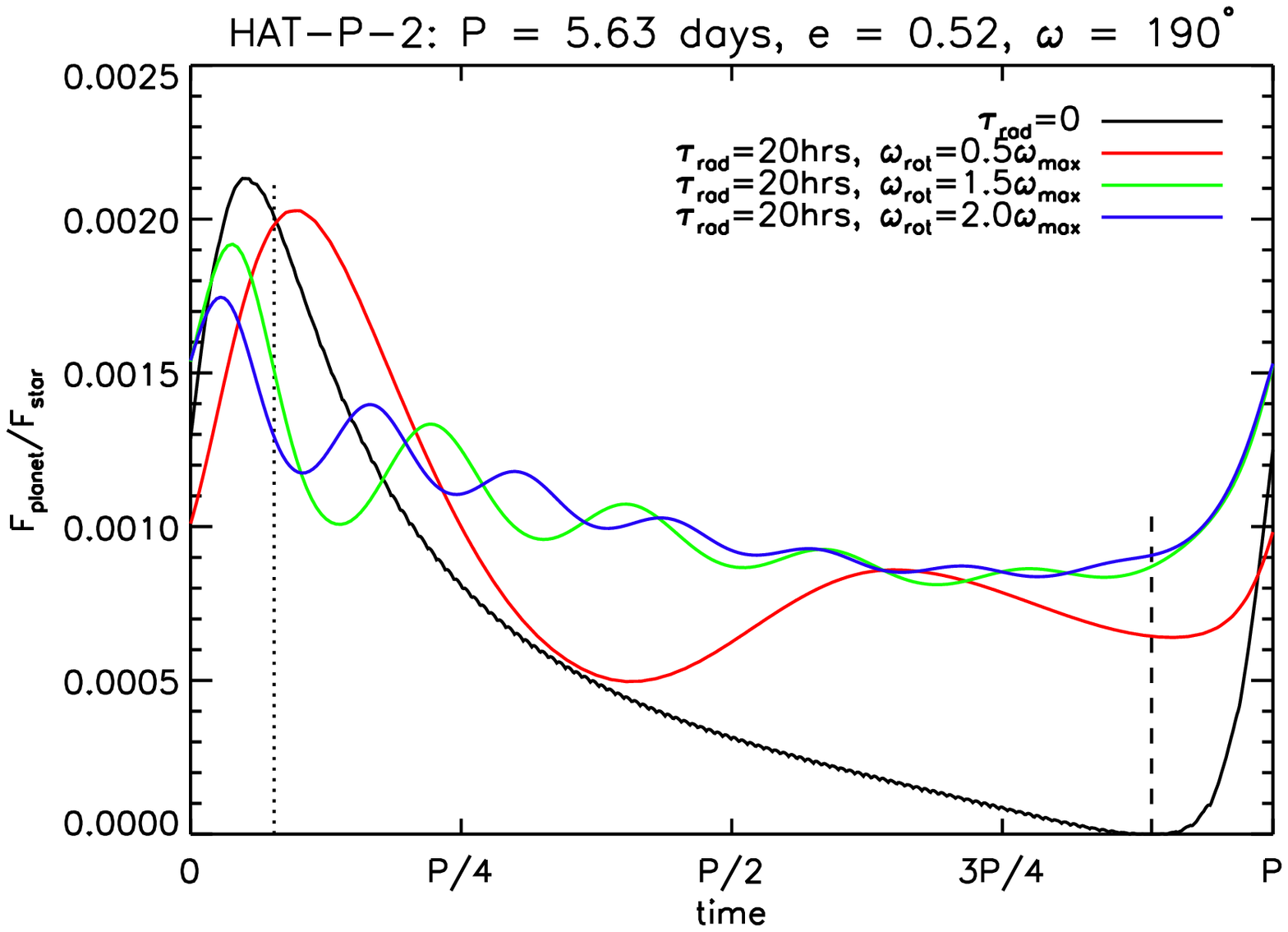}
\caption{The 8~$\mu$m light curve for HAT-P-2b \citep{Bakos_2007} for a variety of radiative and advective timescales, and assuming $A=0$. The vertical dotted and dashed lines mark superior conjunction (eclipse) and inferior conjunction (transit), respectively. The transit and eclipses have been removed for clarity.}
\label{sample_HAT-P-2}
\end{figure}

\begin{figure}[tb]
\includegraphics[width=84mm]{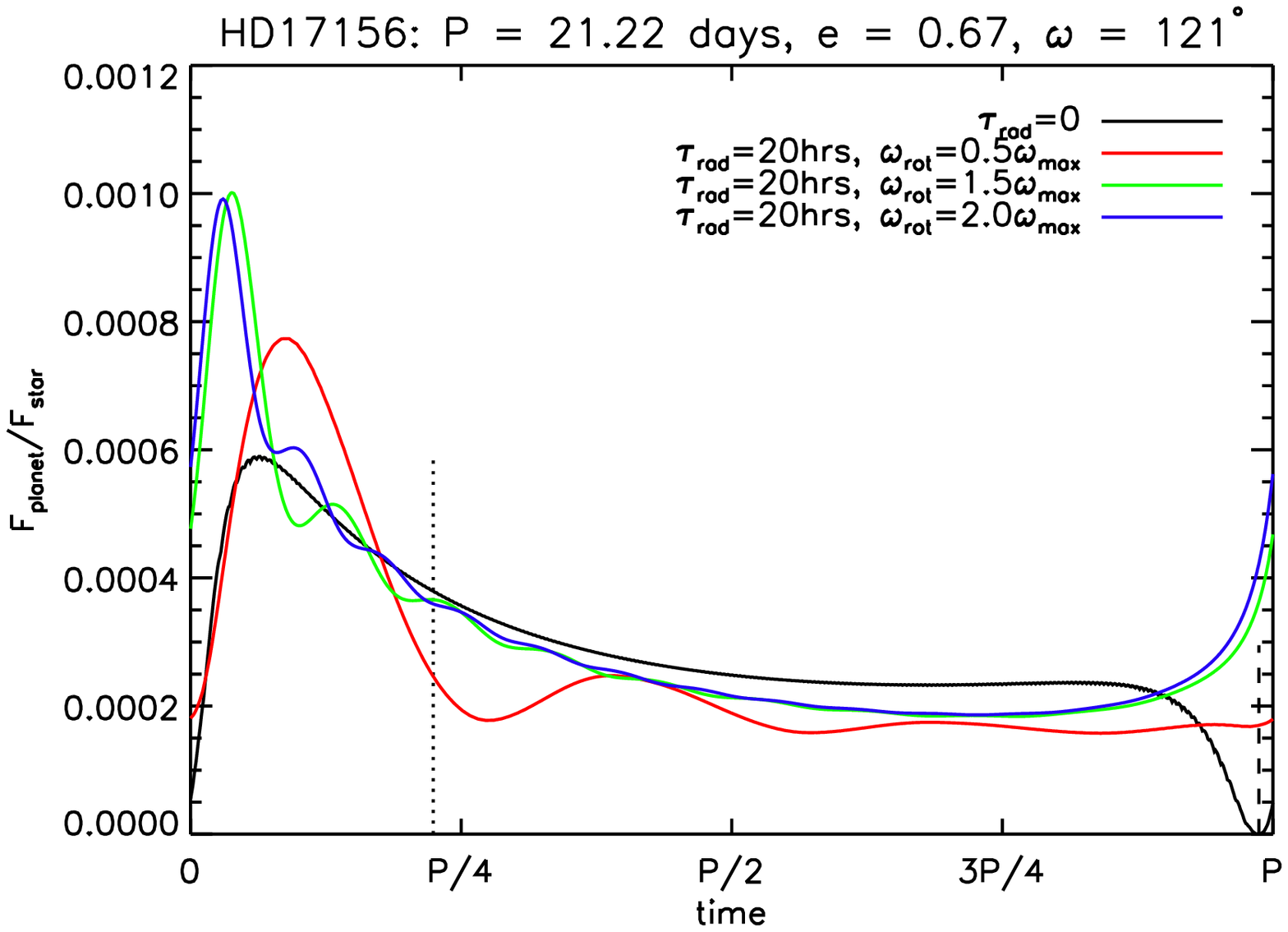}
\caption{The 8~$\mu$m light curve for HD~17156b \citep{Fischer_2007} for a variety of radiative and advective timescales, and assuming $A=0$. The vertical dotted and dashed lines mark superior conjunction (eclipse) and inferior conjunction (transit), respectively. The transit and eclipses have been removed for clarity.}
\label{sample_HD17156}
\end{figure}

\begin{figure}[tb]
\includegraphics[width=84mm]{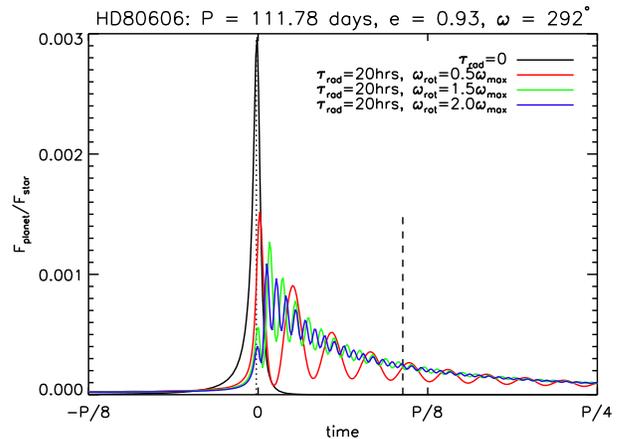}
\caption{The 8~$\mu$m light curve for HD~80606b \citep{Naef_2001} for a variety of radiative and advective timescales, and assuming $A=0$. The vertical dotted and dashed lines mark superior conjunction (eclipse) and inferior conjunction (transit), respectively. The transit and eclipses have been removed for clarity. Models run with $3\times$ higher time-resolution yields indistinguishable light curves.}
\label{sample_HD80606}
\end{figure}

Figures~\ref{sample_WASP-17}--\ref{sample_HD80606} show the same sample of model phase variations, but for planets on eccentric orbits. We use the same $\tau_{\rm rad}$ (0, 20~hrs) and $\omega_{\rm rot}/\omega_{\rm max}$ (0.5, 1.5, 2.0) as for HD~189733b. For consistency and ease of comparison, all planets pass through periastron at $t=0$, and 8~$\mu$m light curves are shown. In many cases \emph{Spitzer} light curves have been obtained at this wavelength, and since we use a single-layer model the shapes of the modelled lightcurves would be similar at other wavelengths. Furthermore, we assume $A=0$ across the board. Non-zero albedo would reduce the mean planet/star thermal contrast, but not the \emph{shape} of the light curves. Furthermore, light curves at multiple wavelengths are necessary to break the degeneracy between albedo and wavelength-dependence: a small DC-offset in a single waveband either indicates that the planet has reflected away a fraction of incident flux, or that the absorbed energy is being re-radiated at other wavelengths.  

\subsection{Discussion}
The black lines (the $\tau_{\rm rad}=0$ case) can be understood purely geometrically: the minimum contrast ratio of zero occurs at transit (marked with a vertical dashed line), since at that phase only the planet's night side is visible. The maximum contrast ratio ---on the other hand--- does not necessarily coincide with superior conjunction (the vertical dotted line) for non-zero eccentricity. If the incident flux is increasing at superior conjunction ($|\omega|<90^{\circ}$), the maximum in the light curve occurs somewhat late (after eclipse); if the incident flux is decreasing at superior conjunction ($90^{\circ}<\omega<270^{\circ}$), the peak occurs early (before superior conjunction). 

Including advection complicates this picture: super-rotating and trailing winds have the effect of shifting the peak early and late, respectively (see first Figure~\ref{sample_HD189733}). The direction of the winds at periastron (whether $\omega_{\rm rot}$ is greater than or less than $\omega_{\rm max}$) determines the offset of the thermal phase peak from the $\tau_{\rm rad}=0$ case. 

We present the planets in order of increasing eccentricity, which also roughly corresponds to increasing orbital period. WASP-17b (Figure~\ref{sample_WASP-17}) has a very ``circular'' phase variation in the $\tau_{\rm rad}=0$ case, but the symmetry between the $\omega_{\rm rot}/\omega_{\rm max} = 0.5$ and 1.5 cases is broken for non-zero radiative time. 

Gl~436b, HAT-P-11b and XO-3b (Figures~\ref{sample_Gl436}, \ref{sample_HAT-P-11} and \ref{sample_XO-3}) show how the argument of periastron, $\omega$, is critical in determining the shape of the phase variations: the periods and eccentricities of the three planets differ, but they all have periastron shortly after superior conjunction and therefore strikingly similar light curves. For comparison, \cite{Deming_2007} and \cite{Demory_2007} measured 8~$\mu$m secondary eclipse depths of 5.7(8)$\times10^{-4}$ and 5.4(7)$\times10^{-4}$, respectively, for Gl~436b. The multi-band eclipse depths of \cite{Stevenson_2010} indicate the planet has a hot day-side; this is most consistent with the $50\times$ solar metalicity model of \cite{Lewis_2010}, which exhibits a super-rotating equatorial jet. The 8~micron model phase curves presented by \cite{Lewis_2010} are similar to our super-rotating solutions (the blue and green curves in Figure~\ref{sample_Gl436}), and also undershoot the observed eclipse depths. 

HAT-P-2b, HD~17156b and HD~80606b (Figures~\ref{sample_HAT-P-2}, \ref{sample_HD17156} and \ref{sample_HD80606}) have such high eccentricities that even their $\tau_{\rm rad}=0$ lightcurves look nothing like that of a circular planet. Furthermore, their orbital periods are sufficiently long that ---for non-zero $\tau_{\rm rad}$--- they exhibit ``ringing'' as the hot spot generated at periastron rotates in and out of view ($1/\omega_{\rm adv}<\tau_{\rm rad}<P$). The envelope is $\sim\exp((t-t_{\rm peri})/\tau_{\rm rad})$, and the period of the oscillations is simply the planet's effective rotation period, $P_{\rm rot} = 2\pi/\omega_{\rm rot}$. In effect, the ringing caused by flash-heating is not dissimilar to the variability seen in some hydrodynamic simulations due to precession of polar vortices \citep[][]{Cho_2003, Cho_2008}.  

For comparison, \cite{Langton_2008} predicts an 8~$\mu$m phase variation $F_{\rm max}/F_{\rm min}$ of 2.8 for HAT-P-2b, similar to our values of $\sim 2$ for super-rotating winds (see Table~\ref{prediction_table} for a summary of our results).  For HD~17156b, \cite{Irwin_2008} predicts $F_{\rm max}/F_{\rm min}$ of 2.9, while \cite{Iro_2010} predicts 3.3; our models predict somewhat greater phase variations, $F_{\rm max}/F_{\rm min}\sim 5$. Note that if periastron occurs near inferior conjunction, our model predicts that the maximum contrast ratio may actually be \emph{greater} for non-zero $\tau_{\rm rad}$ (e.g., Figure~\ref{sample_HD17156}).

Observations of HD~80606b show an 8~$\mu$m eclipse depth of 1.0(2)$\times10^{-3}$ \citep{Laughlin_2009}, similar to our sub-rotating model (red line in Figure~\ref{sample_HD80606}). For comparison, \cite{Langton_2008} predicts phase variations with $F_{\rm max}/F_{\rm min}=2.1$, \emph{much} smaller that our values of $\sim 60$. Note that they have $\tau_{\rm rad}=4.5(2)$~hours at the 8~$\mu$m photosphere, but the planet never cools to below a contrast ratio of $\sim 4\times 10^{-4}$, despite the planet's 112 day orbit. This discrepancy is largely due to our implicit assumption of no internal heat: after spending over 100 days far from its host star, our model planet cools to $\sim 400$~K. This is still much hotter than the remnant heat of formation (50--100~K), but cooler than the 720~K night-side equilibrium temperature used for the hydrodynamic simulations of \cite{Langton_2008}. Meanwhile, \cite{Iro_2010} predicts $F_{\rm max}/F_{\rm min}=4.2$. The peak to trough amplitudes of our models are within a factor of 2 of the \cite{Langton_2008} and \cite{Iro_2010} simulations. If we adopt the $\tau_{\rm rad}=4.5(2)$~hour of \cite{Langton_2008}, we obtain peak-to-trough amplitudes of $\sim3\times 10^{-3}$, still a factor of $\sim 2$ larger than their models. That factor of 2 is significant: it indicates that ---unlike the \cite{Langton_2008} and \cite{Iro_2010} simulations--- in our model the planet can completely radiate away it's heat over a single planetary orbit.

In general, non-zero heat capacity reduces the maximum contrast ratio and increases the minimum, making for a flatter light curve. Furthermore, the phase amplitude generally diminishes for larger advective frequencies: the peak-to-trough amplitude is smaller for $\omega_{\rm rot} = 1.5\omega_{\rm max}$ than for $\omega_{\rm rot} = 2\omega_{\rm max}$ across the board.  For planets with $90^{\circ}<\omega<270^{\circ}$, however, a second order effect may also take place: the light curve maximum \emph{increases} for moderate $\tau_{\rm rad}$ as this allows the planet to hold onto absorbed power until its hot-spot has rotated into our view (see Figure~\ref{sample_HD17156}).

\section{Model Assumptions} \label{model_assumptions}
\underline{Agnostic about Tidal Locking:} Short-period exoplanets on circular orbits should have tidally locked cores, because the tidal locking timescale is considerably shorter than the age of most planetary systems \citep[e.g.,][]{Lubow_1997}. By the same token, planets on eccentric orbits should have cores in pseudo-synchronous rotation (roughly speaking, this means being tidally locked at periastron, when the tidal forces are strongest). Unlike detailed hydro simulations that use the planet's rotation rate as an input parameter, our model is agnostic about precise prescriptions for pseudo-synchronous rotation frequency, $\omega_{\rm ps}$ \citep[e.g.][]{Hut_1981, Ivanov_2007}. For example, at a moderate eccentricity of $e=0.3$, the \cite{Hut_1981} formulation gives $\omega_{\rm ps} \approx 0.8 \omega_{\rm max}$ , while the \cite{Ivanov_2007} formulation gives $\omega_{\rm ps} \approx 1.4 \omega_{\rm max}$. These details about the planet's interior are not important for our purposes, however, because the planets in our study have thick gaseous envelopes covering their cores and hydrodynamic simulations indicate that wind-speeds in the vicinity of the photosphere are comparable to the rotational velocity. As a result, the parcels of gas are probably not stationary with respect to the sub-stellar point, even for (presumably tidally locked) planets on circular orbits.  As a first-order model we treat them as advecting longitudinally at a constant angular velocity. 

\underline{No Internal Heating:} Tidal circularization is undoubtedly responsible for the generally low eccentricity of short-period planets \citep[e.g.][and references therein]{Ford_2008}.  It is not clear, however, that tidal heating has an observable effect on light curves. For example, \cite{Jackson_2008} calculate present-day tidal heating between $2$ and $5\times 10^{17}$~W for Gl~436b, two orders of magnitude less power than the stellar insolation, which ranges between $7$ and $13\times 10^{19}$~W. For short-period planets, the observational signature of internal heat (either remnant gravitational energy from formation or ongoing tidal heating) will likely be an inflated radius rather than additional emergent flux. If hot Jupiters exist for which the tidal heating is a sizeable fraction of the insolation, then our model will have under-estimated the zeroeth-order (DC) component of their lightcurves. By the same token, we also neglect the remnant heat of formation, which is likely less than 100~K for a Gyrs-old planet \citep[][]{Burrows_2006}. One important consequence of neglecting any internal heating and having a simple one-layer model, is that the night-side equilibrium temperature is zero.  For the most part this is not important, since parcels pass through the illuminated hemisphere often enough to never get very cool.  The exception is for our models of HD~80606b, where the planet spends such a long time ($>100$ days) far from its host star. It is possible that the real planet never cools to the 400~K seen in our toy model.

\underline{No Heat Conduction:} Our model neglects conduction of heat from one cell to its neighbors. As such, our model is only relevant when advection and radiation are the dominant energy transport mechanisms. This is probably the case for hot Jupiters, which are thought to have strong winds. If heat conduction \emph{is} important on hot Jupiters, then they may exhibit smaller day--night temperature contrasts and the relation between the thermal phase offset and the thermal phase amplitude will be broken (see first Equation~12).  

\underline{No Mechanical Work:} Our model assumes that all the incident energy from the host star is either reflected or absorbed, and the energy that is absorbed only goes into increasing the temperature of the parcels of gas. In effect, this amounts to neglecting the work term in the First Law of Thermodynamics. In the steady-state, the power absorbed by the planet must be balanced by the power emitted.  That is to say, the ultimate effect of work is to warm up parcels of gas \citep[e.g., parcels speed up to super-sonic velocities, shock and deposit energy as heat;][]{Goodman_2009}.  This implicitly neglects four effects that may very well be occurring: energy may instead go into latent heat to vaporize particulates; thermal tides \citep{Arras_2009}; increasing the atmospheric scale height \citep{Guillot_2002}; or speeding up winds. Any of these ``work'' terms would tend to dampen the temperature fluctuations experienced by a parcel of gas: they act as energy sinks near periastron and may act as energy sources near apoastron. As such our model will under-estimate gas response time, over-estimate the temperature excursions, and hence the peak-to-trough amplitude of disc-integrated lightcurves.  The planet most likely to show these effects is HD~80606b, since realistically much of the energy absorbed at periastron must go into speeding up winds, inflating the atmosphere, etc. Indeed, the peak-to-trough phase amplitudes we compute for this planet are roughly a factor of 2 greater than those from hydrodynamic simulations \citep{Langton_2008}.

\underline{No Latitudinal Advection:} Our toy model neglects advection towards the planet's poles: effectively we have zonal but no meridional winds. At low (mbar) pressures, there almost certainly is meridional flow away from the sub-stellar point, but the low presures mean that only a small fraction of absorbed power is moved in this way.  But even when meridional flows are not efficient, latitudinal heat transport may still occur in the form of eddy fluxes \citep[e.g.,][]{Cho_2008, Cho_2008b}, which we have also implicitly neglected. Efficient equator--pole heat transport on a planet would have the effect of reducing the DC component of the phase curve. In the limit of perfect latitudinal heat transport, the equatorial temperature is suppressed by a factor $(\pi/4)^{1/4}\approx 0.94$ and the disc-integrated temperature is only suppressed by $(3\pi^{2}/32)^{1/4}\approx 0.98$ compared to the nominal (no latitudinal transport) case. This latitudinal dilution is (necessarily) smaller than the longitudinal dilution. As pointed out in \cite{Budaj_2010}, latitudinal energy transport is only important for exoplanets that are not edge-on (we get a better view of their pole), and in those cases the unknown planetary radius is a confounding factor.

\underline{Single Wind Velocity:} Our model uses a single angular wind velocity, which can roughly be thought of as the velocity of the equatorial jet stream. We are neglecting the jet-streams at mid-latitudes that transport energy in the opposite direction \citep{Dobbs-Dixon_2010}. We have assumed that ---to first order--- the velocity of the equatorial jet is the only speed that affects the observed phase variations of a planet, but the speed of waves may also be important \citep[e.g.,][]{Watkins_2010}. In any case, our $\omega_{\rm rot}$ can be thought of as the visibility-weighted mean of the competing zonal jets, which is necessarily slower than the velocity of the equatorial jet. Note that our adoption of constant \emph{angular} velocity amounts to solid body rotation \citep[see also][]{Iro_2005}. As far as simplifications go, solid-body rotation is attractive because it naturally produces smaller linear velocities near the planet's poles than near the equator, a generic feature of many hydrodynamic simulations.  That being said, the assumption is undoubtedly wrong in detail and one could instead parameterize the zonal flow as having a constant \emph{linear} rather than angular velocity: $\omega_{\rm rot}(\theta) = \omega_{\rm rot} \sin\theta$. Such a parameterization could be normalized to yield the same overall heat transport, but it would smear out thermal phase curve features, since parcels of gas flash-heated at periastron would not move in lockstep across the face of the planet. 

\underline{Constant Wind Velocity:} We assume that the wind speed does not change with time. The existence of steady-state, E-W jet streams is supported by hydrodynamic simulations of planets on circular orbits \citep[][and references therein]{Showman_2009, Rauscher_2009, Dobbs_2008}. But the simulations of \cite{Cho_2003} and \cite{Thrastarson_2010} indicate that these results may be a function of initial conditions, in which case even planets on circular orbits may exhibit wandering equatorial jets. For planets on eccentric orbits, the wind speed of the equatorial jet may very well slow down as the planet's energy budget decreases, but the timing of the primary peak should only depend on the wind speed near pericenter. If the winds damp out near apastron, then the maximum wind velocity may occur shortly after periastron, depending on the inertia of planet's atmosphere. In effect, our model works in the limit of large inertia, so that the winds do not speed up or slow down even as the planet's energy budget changes. There is very little energy to transport when the planet reaches apoastron, so the wind speed at that point in the orbit does not have observable consequences.

\underline{Constant Albedo:} Atmospheric albedo is thought to increase with decreasing temperature.  This can be quite sudden if, for example, clouds form high in the atmosphere \citep[eg,][and references therein]{Kane_2010}. 
As mentioned in \cite{Langton_2008b}, the formation of clouds on an eccentric planet as it moves away from periastron will simply exacerbate its $1/r^{2}$ energy budget, making the periastron passage even more important for thermal light curves. Note that this is the only of our model assumptions that could lead us to \emph{under}-estimate the amplitude of phase variations.

\underline{Blackbody Emitters:} While we treat the parcels of gas as blackbody emitters, the \emph{shapes} of our lightcurves should not change much if the planet has a non-blackbody spectrum. The shape of the light curves is preserved provided that the mid-IR flux increases as a simple function of temperature. This is likely to be true on the Rayleigh-Jeans tail of the planet's spectral energy distribution.

\underline{Vertically Isothermal:} Using a single temperature at each position on the planet (ie: a single-layer model) implicitly means assuming that the temperature at the emitting surface is roughly equal at all wavelengths of interest. In the case of highly irradiated planets, the day-side temperature-pressure profile is shallower than for an isolated giant planet. Depending on how much advection of heat occurs ---and at what depths--- this may not be true of the night-side. Observationally, the day-side brightness temperatures of hot Jupiters often differ from one infrared waveband to another.  But the only hot Jupiter with thermal phase curves observed at multiple wavelengths is HD~189733b, and the \emph{shape} of the 8 and 24 micron phase curves are indistinguishable, given the uncertainties \citep{Knutson_2009a}.  This is notable since 1-D atmospheric models of this planet place the 8~micron photosphere at more than twice the pressure of the 24 micron photosphere. Note that this coincidence for HD~189733b only means that the \emph{ratio} of the radiative and advective times is the similar at the 8 and 24 micron photospheres. For eccentric planets the degeneracy between those two timescales is broken, so seeing the same morphology at two different wavebands means that \emph{both} the radiative and advective timescales are the same at the two photospheres. If the phase variation morphology is grossly different from one waveband to another, then one would need a model with different radiative and advective times as a function of waveband (or equivalently, a multi-layer model).  

\section{Conclusions}
We have presented a semi-analytic model for the heating pattern of gas in the atmosphere of a hot Jupiter. Our two physically-motivated parameters are the planet's radiative timescale, and its wind speed. This model is meant to calculate the shape of thermal phase variations of transiting exoplanets with any eccentricity or argument of periastron. We have assumed zero albedo and blackbody-like emission throughout: these assumptions have an effect on the DC offset of our phase variations (e.g., the secondary eclipse depth), but should not significantly affect the shape of the phase variations (timing of the phase peak, peak-to-trough amplitude, etc.). 

\begin{deluxetable}{lrrrr}
\tabletypesize{\scriptsize}
\tablecaption{Calculated Phase Variations of Eccentric Planets \label{prediction_table}}
\tablewidth{0pt}
\tablehead{
\colhead{Planet} & \colhead{$\tau_{\rm rad}$} & \colhead{$\omega_{\rm rot}/\omega_{\rm max}$} & \colhead{Amplitude$^{b}$} & \colhead{$\Delta t_{\rm max}^{c}$}}
\startdata
HD~189733b & 0 hrs & N/A & $4.9\times10^{-3}$& 0.0 hrs\\
& 20 hrs & 0.5 & 2.9& 5.3 hrs\\
& 20 hrs & 1.5 & 2.9& -5.3 hrs\\
& 20 hrs & 2.0 & 1.9& -7.2 hrs\\
WASP-17b & $0$ hrs & N/A & $4.9\times10^{-3}$& 0.27 hrs\\
& 20 hrs & 0.5 & 3.9& 5.8 hrs\\
& 20 hrs & 1.5 & 2.4& -2.4 hrs\\
& 20 hrs & 2.0 & 1.8& -4.7 hrs\\
Gl~436b & $0$ hrs & N/A & $8.0\times10^{-4}$ & 1.8  hrs\\
& 20 hrs & 0.5 & 6.5& 5.0 hrs\\
& 20 hrs & 1.5 & 2.4& -24 hrs\\
& 20 hrs & 2.0 & 1.7& -31 hrs\\
HAT-P-11b & 0 hrs & N/A & $5.1\times10^{-4}$ & 2.6  hrs\\
& 20 hrs& 0.5 & 9.4& 6.7 hrs\\
& 20 hrs& 1.5 & 2.7& -2.2 hrs\\
& 20 hrs& 2.0 & 1.8& -4.7 hrs\\
XO-3b & 0 hrs & N/A & $2.6\times10^{-3}$ & 1.6 hrs\\
& 20 hrs& 0.5 & 3.9& 3.8 hrs\\
& 20 hrs& 1.5 & 1.5& -42 hrs\\
& 20 hrs& 2.0 & 1.4& -48 hrs\\
HAT-P-2b & 0 hrs & N/A & $2.1\times10^{-3}$& -3.9 hrs\\
& 20 hrs& 0.5 & 4.1& 3.0 hrs\\
& 20 hrs& 1.5 & 2.4& -5.1 hrs\\
& 20 hrs& 2.0 & 2.1& -6.6 hrs\\
HD~17156b & 0 hrs & N/A & $5.9\times10^{-4}$& -82 hrs\\
& 20 hrs& 0.5 & 4.9& -68 hrs\\
& 20 hrs& 1.5 & 5.4& -95 hrs\\
& 20 hrs& 2.0 & 5.3& -99 hrs\\
HD~80606b & 0 hrs & N/A & $3.0\times10^{-3}$ &1.8 hrs\\
& 20 hrs& 0.5 & 63.5& 7.2 hrs\\
& 20 hrs& 1.5 & 56.1& 26 hrs\\
& 20 hrs& 2.0 & 51.0& 21 hrs\\
\enddata
\tablenotetext{a}{We adopt $A=0$ for all these calculations. The shape of the phase variations is not affected by albedo, to first order.}
\tablenotetext{b}{For $\tau_{\rm rad}=0$, this refers to the maximum flux ratio, $F_{\rm max}/F_{*}$: the minimum flux ratio is always zero in that case, so the peak-to-trough amplitude is equal to the peak value. For the non-zero $\tau_{\rm rad}$, the amplitude refers to the \emph{ratio} of the maximum and minimum of the light curve, $F_{\rm max}/F_{\rm min}$.}
\tablenotetext{c}{Time offset between the maximum planet/star contrast and superior conjunction. Negative values mean that the maximum occurs \emph{before} eclipse; positive values mean that maximum occurs \emph{after} eclipse.}
\end{deluxetable}

Our model is sufficient to fit the shape of phase variations for HD~189733b, a hot Jupiter on a circular orbit. The radiative and advective times of our ``best fit'' model for that planet are within the range of plausible parameters from detailed hydrodynamic simulations. We bracket these parameters and compute light curves for a selection of eccentric transiting planets. Our calculated 8~$\mu$m light curves are summarized in Table~\ref{prediction_table}.   

Our principal findings from this numerical experiment are:
\begin{enumerate}
\item For planets on circular orbits, the location on the planet of the primary hot spot is intimately related to its amplitude. 
\item The minimum temperature of a parcel occurs at or shortly after dawn, but the minimum in the disc-integrated phase variations occurs when the primary hotspot is on the far side of the planet rather than when the dawn terminator is facing the observer.
\item For a circular orbit, the timing of the phase variation maximum with respect to superior conjunction indicates the direction of the dominant winds, but cannot break the degeneracy between wind speed and radiative time.
\item For eccentric planets ---in the limit of no advection--- the light curve maximum leads or trails superior conjunction depending on whether the planet is receding from or approaching the star at superior conjunction.
\item Planets with non-zero radiative times have their thermal phase peaks offset early for super-rotating winds, and late for sub-rotating winds.
\item For planets with non-zero $\tau_{\rm rad}$, the peak-to-trough phase amplitude generally decreases with increasing $\omega_{\rm rot}$; for planets on circular orbits this can be understood in terms of the dimensionless advective efficiency, $\epsilon$.
\item For planets that are approaching their star at superior conjunction or planets on circular orbits, increasing $\tau_{\rm rad}$ has the effect of reducing the light curve maximum.
\item Planets that are approaching their star at superior conjunction also exhibit stronger phase variations if they have sub-rotating (rather than super-rotating) winds at periastron. This is because the hot spot created at periastron rotates out of view more slowly.
\item For planets receding from their star at superior conjunction, an increased $\tau_{\rm rad}$ may \emph{increase} the light curve maximum.
\item Eccentric planets with orbital periods significantly longer than $\tau_{\rm rad}$ exhibit ``ringing'' whereby the hot spot generated at periastron rotates in and out of view. 
\item The existence of ringing makes it possible to explicitely measure both $\omega_{\rm rot}$ (the frequency of the ringing) and $\tau_{\rm rad}$ (the damping of the ringing).
\end{enumerate} 
As discussed in \S~6, the myriad assumptions made in our model will tend to exaggerate the amplitude of thermal phase variations. This could lead, for example, to over-estimating the radiative time if one uses our models to interpret observed phase variations.
 
An obvious direction for the future is to couple our simple thermodynamic model to a 1-D radiative transfer code \citep[e.g.][]{Iro_2010} to produce time-variable \emph{spectra} of eccentric planets rather than just light curves.  Such a hybrid solution would require far fewer computational resources than full time-variable radiative transfer.

\acknowledgments
N.B.C. acknowledges useful discussions with E.~Rauscher, as well as N.~Kaib for his help with orbital dynamics, and H.M.~Haggard for his help with differential equations. S.L. Hawley and V.S. Meadows contributed useful comments to the manuscript. N.B.C. was supported by the Natural Sciences and Engineering Research Council of Canada. E.A. is supported by a National Science Foundation Career Grant. Support for this work was provided by NASA through an award issued by JPL/Caltech. N.B.C. acknowledges the hospitality of the Harvard-Smithsonian Center for Astrophysics, and the Kavli Institute for Theoretical Physics, where portions of this work were completed. This research was supported in part by the National Science Foundation under Grant No. NSF PHY05-51164.

\section*{Appendix}
In the slow cooling regime ($\epsilon \gg 1$), the changes in temperature 
during each orbit are small, so one can treat Equation~10
perturbatively.  
The trial solution is $\tilde{T} = \tilde{T}_0 + \delta \tilde{T}$, 
where $\tilde{T}_0=\pi^{-1/4}$ is a constant equal to the average
temperature and $\delta \tilde{T}$ contains all of the variability.
Substituting this into Equation~\ref{circular_de} and noting that $\max(\cos\Phi,0)= \frac{1}{2} (\cos\Phi + \vert \cos\Phi \vert)$, we get:
\begin{equation}
{d\tilde{T}\over d\Phi} =
{1 \over \epsilon} 
\left(\frac{1}{2}\left[\cos{\Phi}+\vert \cos{\Phi}\vert \right]-\frac{1}{\pi} - 4\tilde{T}_0^3
\delta \tilde{T}\right),
\end{equation}
where we have only kept terms up to first-order in $\delta \tilde{T}$.

The solution to this equation is
\begin{eqnarray}
\tilde{T}_{\rm day} &=& \frac{3}{4} \tilde{T}_0 + {\gamma \cos{\Phi} + \sin{\Phi}
\over \epsilon (1+\gamma^2)} + {e^{-\gamma \Phi} \over
2\epsilon (1+\gamma^2) \sinh{(\pi\gamma/2)}} \cr
\tilde{T}_{\rm night} &=& \frac{3}{4} \tilde{T}_0
+ {e^{-\gamma (\Phi-\pi)} \over 2\epsilon (1+\gamma^2) \sinh{(\pi\gamma/2)}},\cr
\gamma &=& {4\tilde{T}_0^3 \over \epsilon}.
\end{eqnarray}

Note that the day-side phase peak (middle term) has the right limits:
for large $\epsilon$, the variation is as $\sin{\Phi}$ (slow-cooling regime), 
while for small $\epsilon$, the variations is as $\cos{\Phi}$ (instant-cooling limit).

In the circular case, the longitude of the peak of the temperature, $\Phi_{\rm max}$, goes 
from $0$ for $\epsilon=0$ to $\pi/2$ for $\epsilon \rightarrow \infty$ (but note that in practice the parcel's heating curve becomes flat before the the peak reaches $\pi/2$, as shown in Figure~\ref{phi_max}).
We compute $\Phi_{\rm max}$ numerically, and fit $\tan(\Phi_{\rm max}(\epsilon))$ with the 
function
\begin{equation}
f(\epsilon)=x_0(1+x_1\epsilon^{-x_2+(x_3/(1+x_4\epsilon))})^{-1},
\end{equation}
for which we find a good fit for 
$x_0=2.9685$, $x_1=7.0623$, $x_2=1.1756$, $x_3=-0.2958$, and $x_4 = 0.1846$.  This
then gives the maximum temperature as a function of $\epsilon$ as 
\begin{equation}
\label{eqn01}
\tilde{T}_{\rm max}(\epsilon) = \cos^{1/4}{\Phi_{\rm max}(\epsilon)} \approx \cos^{1/4}{\left(\tan^{-1}{f(\epsilon)}\right)}.
\end{equation}
The maximum temperature goes from $1$ at small $\epsilon$ (instant re-radiation) to
$\pi^{-1/4}$ at large $\epsilon$ (uniform temperature).
We have computed $T_{\rm max}$ numerically, shown with the asterixes in Figure~\ref{t_sunset_vs_epsilon},
while Equation \ref{eqn01} is plotted as the solid line;  as can be seen the
agreement is quite good.

The next critical characteristics are the temperatures at dawn and sunset --- these determine
the temperature of the atmosphere at the terminator, where it absorbs starlight during transit.
We can estimate the sunset temperature as follows.  For very small values of $\epsilon$, 
the flux absorbed on the day-side is emitted nearly instantly, so heating balances cooling, 
and $\tilde{T}_{\rm day}(\epsilon \ll 1) \approx \cos^{1/4} \Phi$.
However, near sunset, $\Phi=\pi/2$, the temperature drop becomes so significant
that $d\tilde{T}_{\rm day}/d\Phi$ becomes comparable to the cooling rate, $-\cos{\Phi}$.
At this point the planet's temperature is no longer determined by instant re-radiation 
of the absorbed heat; instead we need to take into account the cooling timescale.
For small $\epsilon$ this occurs very close to $\pi/2$, so we can estimate the temperature at 
sunset as the time when $\epsilon d\tilde{T}_{\rm day}(\epsilon \ll 1) \approx -\cos{\Phi}$.
This gives $\epsilon d[\cos^{1/4}{\Phi}]/d\Phi = -\frac{1}{4}\epsilon \cos^{-3/4}{\Phi} \sin{\Phi}
\approx -\cos{\Phi}$.  Since $\Phi \approx \pi/2$ and $\tilde{T} \approx \cos^{1/4} \Phi$,
this gives $\tilde{T}_{\rm dusk} \approx (\epsilon/4)^{1/7}$.  This is a very weak dependence on
$\epsilon$.  For large values of $\epsilon$, recirculation is very efficient, so
$\tilde{T}_{\rm dusk} \approx \pi^{-1/4}$.  Combining these two limits, we find the
approximate formula $\tilde{T}_{\rm dusk} \approx [\pi^2(1+y_0/\epsilon)^{-8}+y_1\epsilon^{-8/7}]^{-1/8}$,
where $y_0 = 0.69073$ and $y_1 = 7.5534$, which is plotted as the dotted line in
Figure~\ref{t_sunset_vs_epsilon}, as well as numerically computed data points (diamonds); again the agreement
is quite good.  

Finally, the minimum in temperature occurs at (or very close to) the dawn terminator 
after the gas has cooled all night and before it starts warming again on the day side.  
The dawn temperature is well approximated by $\tilde{T}_{\rm dawn} = 
[\pi+(3\pi/\epsilon)^{4/3}]^{-1/4}$, plotted as the dashed line in Figure~\ref{t_sunset_vs_epsilon},
as well as numerically computed data points (triangles).  The agreement for this expression
is quite accurate as well.

\end{document}